# Evidence of 100 TeV $\gamma$-ray emission from HESS J1702-420: A new PeVatron candidate

**L. Giunti,**[a,*] **B. Khélifi,**[a] **K. Kosack,**[b] **R. Térrier**[a] **for the H.E.S.S. Collaboration**
(a complete list of authors can be found at the end of the proceedings)

[a] *Université de Paris, CNRS, Laboratoire Astroparticule et Cosmologie, F-75013 Paris, France*
[b] *IRFU, CEA, Université Paris-Saclay, F-91191 Gif-sur-Yvette, France*
*Email:* contact.hess@hess-experiment.eu

The identification of active PeVatrons, hadronic particle accelerators reaching the knee of the cosmic-ray spectrum (at the energy of few PeV), is crucial to understand the origin of cosmic rays in the Galaxy. In this context, we report on new H.E.S.S. observations of the PeVatron candidate HESS J1702-420, which bring evidence for the presence of $\gamma$-rays up to 100 TeV. This is the first time in the history of H.E.S.S. that photons with such high energy are observed. Remarkably, the new deep observations allowed the discovery of a new $\gamma$-ray source component, called HESS J1702-420A, that was previously hidden under the bulk emission traditionally associated with HESS J1702-420. This new object has a power-law spectral slope $< 2$ and a $\gamma$-ray spectrum that, extending with no sign of curvature up to 100 TeV, makes it an excellent candidate site for the presence of PeV-energy cosmic rays. This discovery brings new information to the ongoing debate on the nature of the unidentified source HESSJ1702-420, and on the origin of Galactic cosmic rays.

*Speaker

## 1. Introduction

The acceleration sites of cosmic rays are a century-old unknown in modern astrophysics. The current understanding is that the bulk of cosmic rays reaching Earth — mostly energetic protons — originate within our Galaxy, outside of the solar system, at unknown sites where they are accelerated up to the energy of the knee feature in the cosmic ray spectrum. Since the measured knee energy is around $3-4$ PeV (1), the Galactic accelerators responsible for cosmic rays up to the knee are called PeVatrons. Several source populations, such as supernova remnants (SNRs) and young massive stellar clusters, have been proposed as potential PeVatron candidates, but to date no observation has definitively linked any particular source class to the acceleration of PeV protons. The H.E.S.S. Collaboration has already reported evidence for the acceleration of PeV protons in the central molecular zone around Sgr A$^*$(2; 3), at a level that is presently insufficient to sustain the flux of PeV cosmic rays observed at Earth. Recently, LHAASO has detected 12 sources in the Northern $\gamma$-ray sky, at energies $>100$ TeV (4). Apart for the Crab Nebula, the nature of the primary accelerators powering those sources is still unclear. HESS J1702-420 is a VHE $\gamma$-ray source without known multi-wavelength counterpart, discovered during the first H.E.S.S. Galactic plane survey campaign in the Southern sky with a significance of $4\sigma$, based on a $5.7$ hr observation livetime (5). In (6), a dedicated analysis revealed a hard power law spectral index of $\Gamma = 2.07 \pm 0.08_{\rm stat} \pm 0.20_{\rm sys}$, with no sign of cut-off, and a significantly extended morphology which is well described by a $0.30^{\circ} \times 0.15^{\circ}$ elliptical Gaussian template. With better reconstruction and data selection algorithms, the HGPS catalog (7) confirmed the spectral hardness of the source, $\Gamma = 2.09 \pm 0.07_{\rm stat} \pm 0.20_{\rm sys}$, and estimated a source significance of $15\sigma$ based on 9.5 hr of observations.

We report on new H.E.S.S. observations of HESS J1702-420 that have been processed with improved techniques (8). Additionally, archival *Fermi*-LAT data were analyzed, to perform a broadband modeling of the TeV source.

## 2. H.E.S.S. data analysis and results

The results presented in this paper make use of data collected from 2004 to 2019, using observations from the CT1-4 H.E.S.S. array for a total of $44.9$ hr acceptance-corrected livetime on the source. Observations were processed using the H.E.S.S. analysis package (HAP), with a dedicated configuration which maximizes the collection area at high energies ($E > 1$ TeV). The reduced data together with the instrument response functions (IRFs) [1] were exported to FITS files, then all high-level analysis results were obtained using `gammapy` (version 0.17), an open source python package for $\gamma$-ray astronomy (9; 10; 11). We performed a three-dimensional (3D) binned likelihood analysis in a $4^{\circ} \times 4^{\circ}$ region of interest (RoI) encompassing HESS J1702-420, above 2 TeV. This technique, recently introduced in the VHE $\gamma$-ray astronomy domain (12; 13), allows to adjust a parametric spectro-morphological model to a data cube, which carries information on the number of reconstructed events within each energy and spatial bin. The optimal source model for the RoI was determined using a statistical approach based on the iterative addition of new components.

[1] They are the effective area, exposure livetime, point-spread function, energy dispersion and field-of-view background model.

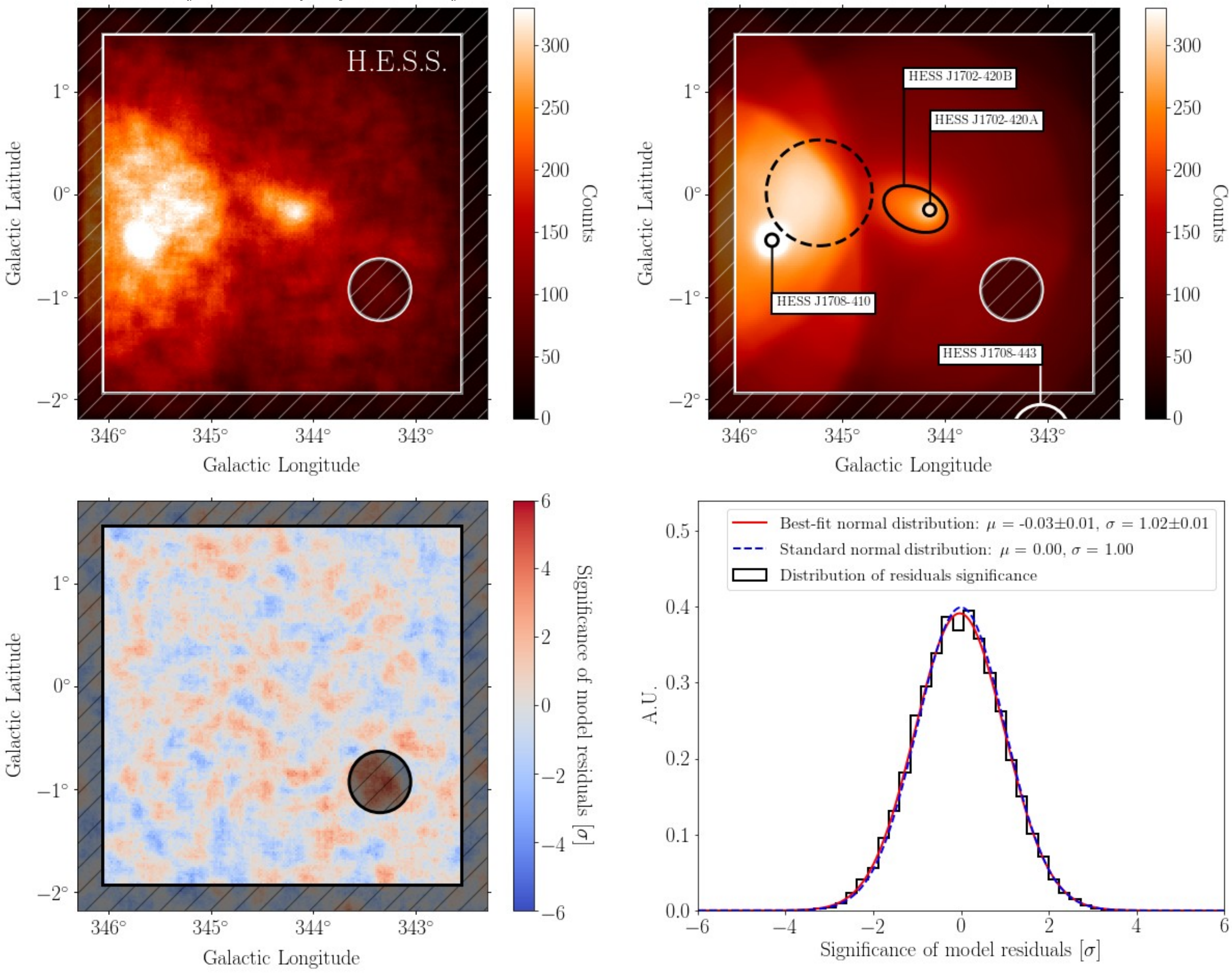

Figure 1: *Upper left panel*: Counts map of the RoI ($E > 2$ TeV), correlated with a $0.1°$-radius top-hat kernel. The hatched regions were excluded from the likelihood computation. *Upper right panel*: Model-predicted counts, with model components overlaid. *Lower left panel*: Spatial distribution of model residuals, showing the statistical significance — in units of Gaussian standard deviations — of *counts - model* fluctuations. *Lower right panel*: Histogram of the significance values from the lower left panel. The adjustment of a Gaussian function to the histogram is shown, together with a reference standard normal distribution.

Step-by-step, the improvement of the source model was assessed by using the likelihood-ratio test and visual inspecting the flattening of spatial and spectral residuals toward zero.

The result of the 3D analysis is shown in Figure 1. The top left (right) panel shows the measured (model-predicted) counts map. Diagonal line hatches represent portions of the RoI that were excluded from the likelihood computation, to limit edge effects and remove an unmodeled $\approx 3\sigma$ hotspot. The measured data are well matched by the model prediction, since the spatial distribution of the significance of model residuals (bottom left panel) does not contain significant structures. The histogram of significance values (bottom right panel) closely follows a standard normal distribution, as expected if residuals are only due to statistical Poisson fluctuations. The top right panel of Figure 1 also shows the $1\sigma$ contours of all components found in the final source model. There are two overlapping objects, called HESS J1702-420A and HESS J1702-420B, that together describe the emission from HESS J1702-420. The other known sources HESS J1708-410 and HESS J1708-443 are consistent with the HGPS (7). We also found a large-scale component, indicated by the dashed circle in Figure 1 (top right panel), whose presence was not confirmed by the crosscheck analysis. Future observation will help to assess its existence and nature.

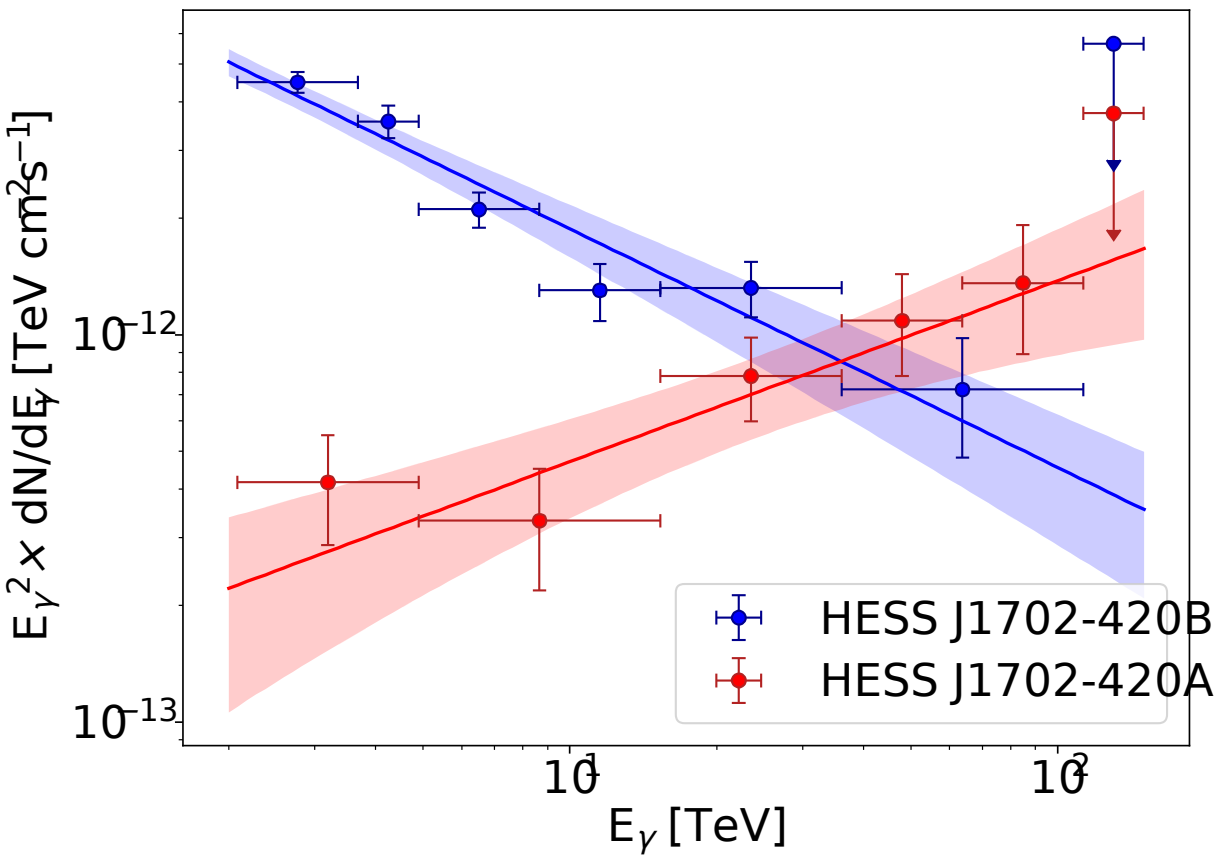

Figure 2: Power law spectra of HESS J1702-420A (red solid line) and HESS J1702-420B (blue solid line), as a function of the incident photon energy $E_\gamma$. The butterfly envelopes indicate the 1 $\sigma$ statistical uncertainty on the spectral shape.

The most relevant result for the identification of Galactic Pevatrons, as well as determining their nature, is the discovery — with a TS-based confidence level corresponding to $5.4\sigma$ — of a new source component, HESS J1702-420A, hidden under the bulk emission formerly associated with HESS J1702-420. This object has a spectral index of $\Gamma = 1.53 \pm 0.19_{\rm stat} \pm 0.20_{\rm sys}$ and a $\gamma$-ray spectrum that, extending with no sign of curvature up to at least 64 TeV (possibly 100 TeV), makes it a compelling candidate site for the presence of extremely high energy cosmic rays. With a flux of $(2.08 \pm 0.49_{\rm stat} \pm 0.62_{\rm sys}) \times 10^{-13}\,{\rm cm}^{-2}\,{\rm s}^{-1}$ above 2 TeV and a $1\sigma$ radius of $(0.06 \pm 0.02_{\rm stat} \pm 0.03_{\rm sys})^\circ$, HESS J1702-420A is outshone below $\approx 40$ TeV by the companion HESS J1702-420B. The test of a point-source hypothesis for HESS J1702-420A resulted in a non-convergence of the fit. HESS J1702-420B has a steep spectral index of $\Gamma = 2.62 \pm 0.10_{\rm stat} \pm 0.20_{\rm sys}$, elongated shape and a flux above 2 TeV of $(1.57 \pm 0.12_{\rm stat} \pm 0.47_{\rm sys})\,10^{-12}\,{\rm cm}^{-2}\,{\rm s}^{-1}$ that accounts for most of the low-energy HESS J1702-420 emission. For neither of the two sources did an exponential cut-off function statistically improve the fit with respect to a simple power law (cut-off significance $1\sigma$). The $\gamma$-ray spectra of both components are shown in Figure 2, together with spectral points obtained by rescaling the amplitude of the reference power law in each energy bin. HESS J1702-420B is the brightest component up until roughly 40 TeV, where HESS J1702-420A eventually starts dominating with its $\Gamma \approx 1.5$ power law spectrum up to 100 TeV. The second to last spectral point of HESS J1702-420A (HESS J1702-420B), covering the reconstructed $64-113$ TeV ($36-113$ TeV) range, has a $4.0\sigma$ ($3.2\sigma$) significance.

As a complementary study, we used the adaptive ring background estimation method (14) to measure the $\gamma$-ray flux above 2, 5, 15 and 40 TeV inside a $1.6^\circ \times 1.6^\circ$ region encompassing HESS J1702-420 (see Figure 3). The figure suggests a shrinking of the VHE emission at high energy, with a shift of the $\gamma$-ray peak toward the position of the unidentified source *Suzaku* src B. Based on the 3D analysis results, this effect is understood as the transition between a low energy regime — dominated by the steep spectrum of HESS J1702-420B — to a high energy one, in which HESS J1702-420A stands out with its exceptionally hard power law spectrum.

## 3. Discussion

Owing to the `NaimaSpectralModel` class implemented in `gammapy`, we could forward-fold the physically-motivated `naima` radiative models directly on the H.E.S.S. 3D data. This represents a significant improvement on the typical `naima` fit to precomputed flux points, which is inevitably biased by the spectral assumption underlying the flux point computation. The (unknown) source distance and target gas density values were chosen arbitrarily, being both degenerate with the source intrinsic luminosity. In the leptonic scenario, we considered inverse Compton up-scattering of the CMB and infra-red radiation fields from (15).

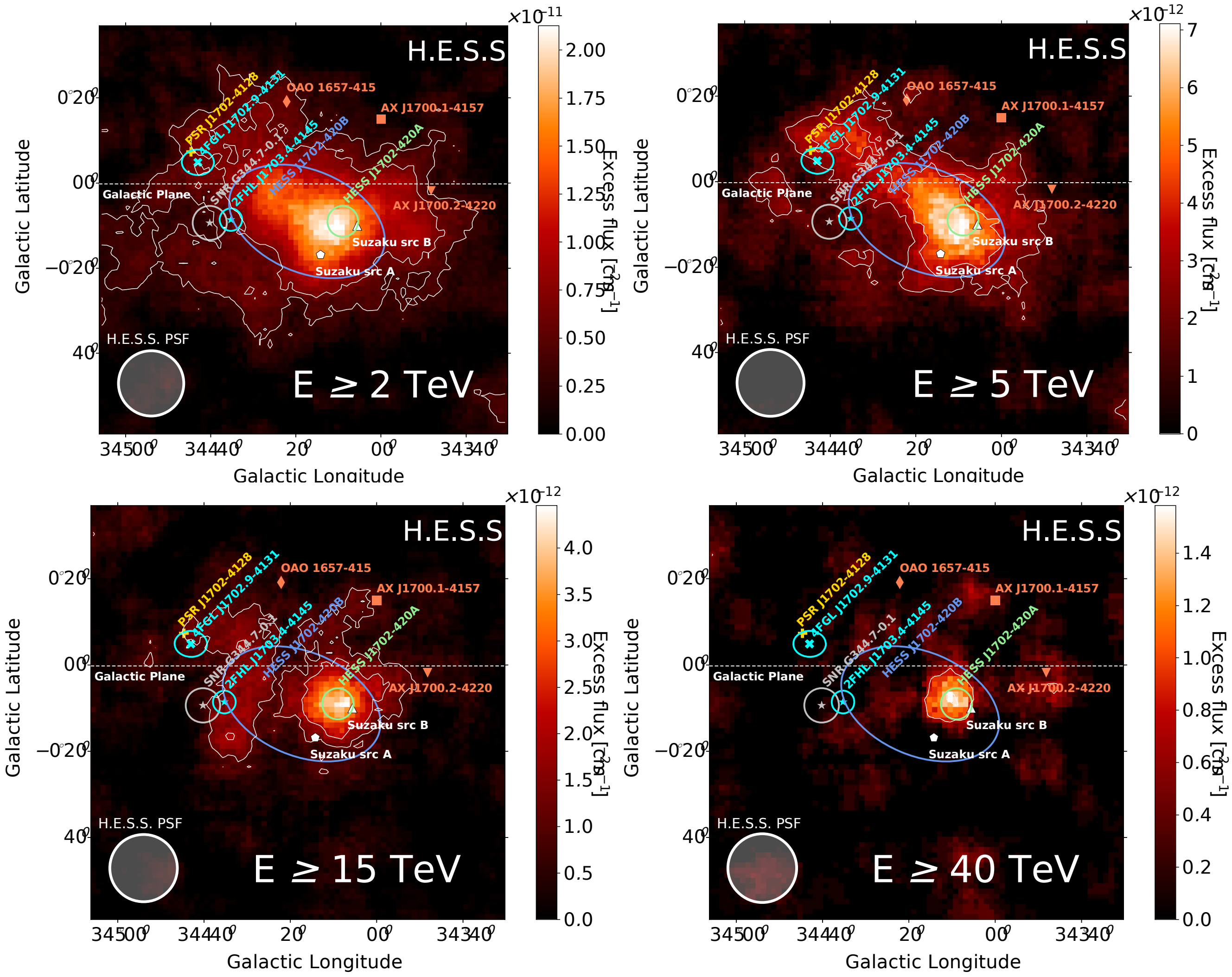

Figure 3: $\gamma$-ray flux maps of the HESS J1702-420 region, computed with the Ring Background Method, above 2 (*top left*), 5 (*top right*), 15 (*bottom left*) and 40 (*bottom right*) TeV. All maps are correlated with a $0.1^{\circ}$-radius top-hat kernel, and the color code is in unit of $\gamma$-ray flux (cm$^{-2}$ s$^{-1}$) per smoothing area. The white contours indicate the 3 $\sigma$ and 5 $\sigma$ H.E.S.S. significance levels. The positions of known astrophysical objects, together with the H.E.S.S. PSF, are also shown in each panel.

A pure power law distribution of protons (electrons) with slope $\Gamma_p = 1.58 \pm 0.14_{\mathrm{stat}}$ ($\Gamma_e = 1.61 \pm 0.15_{\mathrm{stat}}$) well describes the $\gamma$-ray emission of HESS J1702-420A, via hadronic (leptonic) radiative processes. The two spectra, with their 1 $\sigma$ butterfly envelopes, are shown in Figure 4 (left panel). We computed lower limits on the particle cut-off energy, using a Gaussian prior on the particle spectral index to prevent it from floating toward nonphysical regions (i.e., very small or even negative values). In the case of the hadronic model, we assumed as a prior a Gaussian distribution centered at $\Gamma_p = 2$ and with $\sigma = 0.5$, based on standard diffusive shock acceleration theory (DSA; (16)). We found that for a prior centered at $\Gamma_p = 2$ (1.7, 2.3) the 95% confidence-level lower limit on the proton cut-off energy is 0.82 (0.55, 1.16) PeV. The fact that — independenly of the chosen prior — the cut-off energy lower limit is found at $E_p > 0.5$ PeV means that in a hadronic scenario the source likely harbors PeV cosmic rays. In a leptonic scenario instead, assuming $\Gamma_e =$ 2.0 (1.5, 2.5), the 95% confidence-level lower limit on the electron cut-off energy is 106 (64, 152) TeV.The energy contents in protons and electrons, necessary to sustain the $\gamma$-ray emission of HESS J1702-420A, are $W_p(E_p > 1\text{TeV}) \gtrsim 1.8 \times 10^{47} (d/3.5 \text{ kpc})^2 \, (n_\text{H}/100 \text{ cm}^{-3})^{-1}$ erg and $W_e(E_e > 1\text{TeV}) \gtrsim 8.1 \times 10^{45} (d/3.5 \text{ kpc})^2$ erg , respectively.

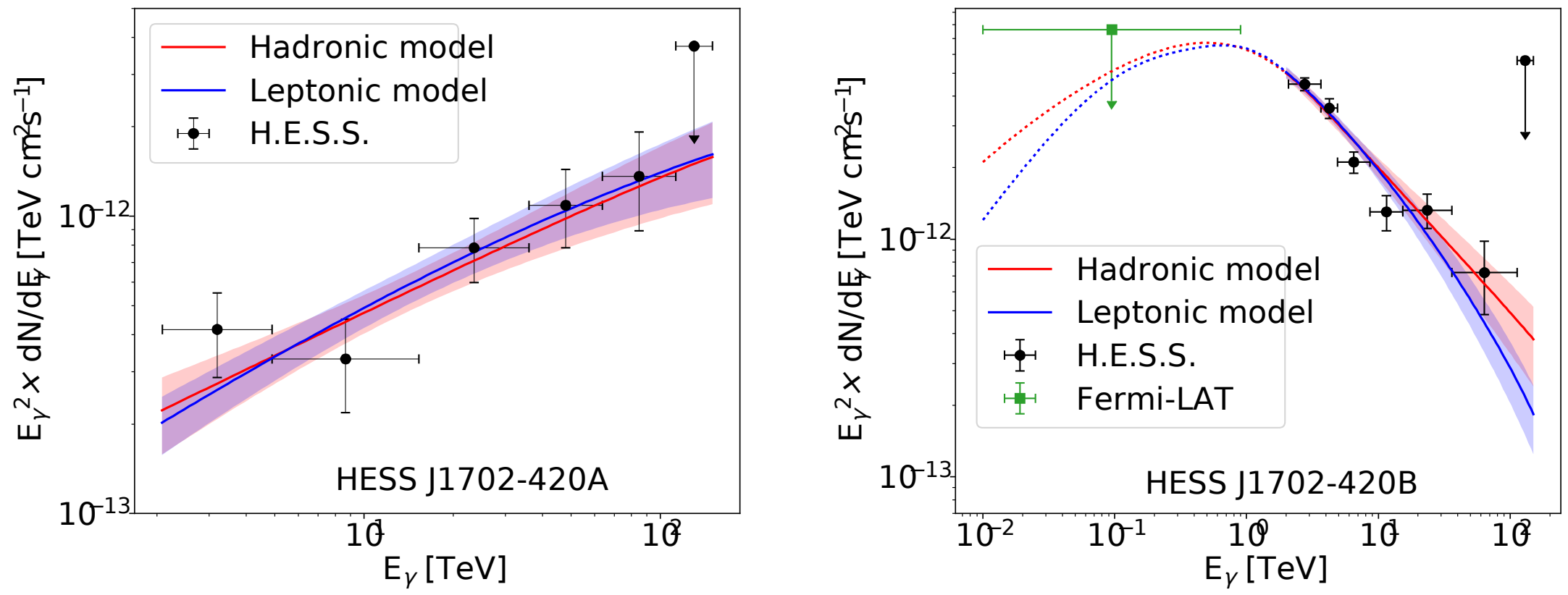

Figure 4: Models of $\gamma$-ray emission based on hadronic (red) and leptonic (blue) one-zone scenarios, for HESS J1702-420A — *left* panel — and HESS J1702-420B — *right* panel. The best-fit spectra, under the assumption of simple power law distribution of the underlying particle populations, are shown as solid lines, while the shaded areas and dotted lines represent the 1 $\sigma$ statistical error envelope and extrapolations outside the fit range, respectively. The H.E.S.S. and Fermi-LAT flux points are also shown, for reference purpose.

In a leptonic scenario, HESS J1702-420A would be powered by an electron population with unusually hard spectral index, $\Gamma_e \approx 1.6$ , and the electron energy required to power the $\gamma$-ray emission would be high compared to the typical values for TeV detected PWNe (17). A simple one-zone leptonic model is therefore challenged, also because it would imply the presence of inverse-Compton emitting electrons with $E_e \approx 100$ TeV. Indeed, given the $\propto 1/E_e$ dependence of the synchrotron loss timescale in the Thomson regime, such energetic electrons would cool down extremely fast creating a high energy spectral curvature or break, which is not observed for HESS J1702-420A. Also, the only known nearby pulsar is PSR J1702-4128, that to power the whole HESS J1702-420 would require an extremely high conversion efficiency ($\approx 20\%$)[2]. As visible in figure 3, *Suzaku* has detected a faint X-ray source called src B positionally close to HESS J1702-420A (18). Based on its measured X-ray flux [3], our simplistic one-zone leptonic model implies an unrealistical magnetic field value of $B \approx 0.3\,\mu G$, which disfavor this multi-wavelength association. We notice that an alternative interpretation is possible, in which the observed $\gamma$-ray emission is due to electrons that are accelerated by the reconnection electric field at X-points in the current sheets of a pulsar striped wind, where the magnetic field value is expected to be low (19). If true, this would be the first time that a TeV measurement probes the reconnection spectrum immediately downstream of the termination shock of a pulsar wind.

[2]But it remains possible that it powers at least part of HESS J1702-420B. Indeed, significant VHE $\gamma$-ray emission is detected by H.E.S.S. near the pulsar position — see Figure 3 (upper right panel).

[3]Which may however suffer from strong systematics (in particular be underestimated) due to edge effects at the borders of the Suzaku field of view.

In a hadronic scenario, the 100 TeV $\gamma$-ray emission from HESS J1702-420A, together with its proton cut-off energy lower limit at $0.55 - 1.16$ PeV, would make it a compelling candidate site for the presence of PeV cosmic ray protons. Therefore HESS J1702-420A becomes one of the most solid PeVatron candidates detected in H.E.S.S. data, also based on the modest value of the total energy in protons that is necessary to power its $\gamma$-ray emission and the excellent agreement of a simple proton power law spectrum with the data. However, we notice that a proton spectrum with a slope of $\Gamma_p \approx 1.6$ over two energy decades is hard to achieve in the standard DSA framework (16). This fact may suggest that HESS J1702-420A, instead of being a proton accelerator, is in fact a gas cloud that, being illuminated by cosmic rays transported from elsewhere, acts as a passive $\gamma$-ray emitter. In that case, the hard measured proton spectrum could result from the energy-dependent particle escape from a nearby proton PeVatron (20). Alternatively, the $\gamma$-ray emission from HESS J1702-420A might be interpreted as the hard high energy end of a concave spectrum arising from nonlinear DSA effects (21). Also, it might originate from the interaction of SNR shock waves with a young stellar cluster wind (22), or cosmic ray interactions with turbulent plasma near OB Associations (23). The absence of a clear spatial correlation between the ISM and the observed TeV emission (24) prevents however a confirmation of the hadronic scenario, unless an extremely powerful hidden PeVatron is present. In the latter case, even a modest gas density would suffice to produce the measured $\gamma$-ray emission of HESS J1702-420A, which would explain the observed nonlinearity between the ISM and TeV maps.

Finally, the baseline proton and electron spectra used to model HESS J1702-420B are broken power laws (see Figure 4, right panel). The presence of a spectral break may be interpreted as a signature of energy-dependent particle escape. In the hadronic (leptonic) scenario, the best-fit proton (electron) spectrum corresponds to a broken power-law with slopes $\alpha_1 = 1.6$ (1.4) and $\alpha_2 = 2.66 \pm 0.11_{\rm stat}$ ($3.39 \pm 0.11_{\rm stat}$), and with break energy of $\tilde{E} = (6.77 \pm 3.64_{\rm stat})$ TeV ($(4.19 \pm 1.25_{\rm stat})$ TeV ). The values of proton and electron energetics, necessary to power the $\gamma$-ray emission of HESS J1702-420B, are respectively $W_p(E_p > 1\,{\rm GeV}) \approx 2.8 \times 10^{48} \left(\frac{d}{3.5\,{\rm kpc}}\right)^2 \left(\frac{n_{\rm H}}{100\,{\rm cm}^{-3}}\right)^{-1}$ erg and $W_e(E_e > 1\,{\rm GeV}) \approx 4.5 \times 10^{47} \left(\frac{d}{3.5\,{\rm kpc}}\right)^2$ erg .

## 4. Conclusions

We present new H.E.S.S. observations of the unidentified source HESS J1702-420, processed with a high-energy oriented analysis configuration (8). We performed a 3D likelihood analysis with `gammapy`, which allowed to separate for the first time two components — both detected at $> 5\sigma$ confidence level — inside HESS J1702-420. We report on the $4.0\sigma$ confidence level detection of $\gamma$-ray emission from the hardest component, called HESS J1702-420A, in the energy band $64 - 113$ TeV, which is an unprecedented achievement for the H.E.S.S. experiment and brings evidence for the source emission up to 100 TeV. With a spectral index of $\Gamma = 1.53 \pm 0.19_{\rm stat} \pm 0.20_{\rm sys}$, this object is a compelling candidate site for the presence of PeV cosmic rays. We observe that if this source is of the same type as the new unidentified LHAASO objects (4), then its γ-ray spectrum might extend up to hundreds of TeV. To assess that, a Southern ultra-high energy facility such as SWGO will be necessary. On the other hand, the improved angular resolution of the CTA-South array will help constraining the source morphology, and possibly close the debate on the nature of HESS J1702-420. Observations in the X-ray band will also be important, to search for a multi wavelength counterpart of the TeV source, and clarify the relationship between HESS J1702-420A and the unidentified *Suzaku* src B.

**Acknowledgements:** The support of the Namibian authorities and of the University of Namibia in facilitating the construction and operation of H.E.S.S. is gratefully acknowledged, as is the support by the German Ministry for Education and Research (BMBF), the Max Planck Society, the German Research Foundation (DFG), the Helmholtz Association, the Alexander von Humboldt Foundation, the French Ministry of Higher Education, Research and Innovation, the Centre National de la Recherche Scientifique (CNRS/IN2P3 and CNRS/INSU), the Commissariat à l'énergie atomique et aux énergies alternatives (CEA), the U.K. Science and Technology Facilities Council (STFC), the Knut and Alice Wallenberg Foundation, the National Science Centre, Poland grant no. 2016/22/M/ST9/00382, the South African Department of Science and Technology and National Research Foundation, the University of Namibia, the National Commission on Research, Science Technology of Namibia (NCRST), the Austrian Federal Ministry of Education, Science and Research and the Austrian Science Fund (FWF), the Australian Research Council (ARC), the Japan Society for the Promotion of Science and by the University of Amsterdam. We appreciate the excellent work of the technical support staff in Berlin, Zeuthen, Heidelberg, Palaiseau, Paris, Saclay, Tabingen and in Namibia in the construction and operation of the equipment. This work benefitted from services provided by the H.E.S.S. Virtual Organisation, supported by the national resource providers of the EGI Federation.

## Full Authors List: H.E.S.S. Collaboration

H. Abdalla[1], F. Aharonian[2,3,4], F. Ait Benkhali[3], E.O. Angüner[5], C. Arcaro[6], C. Armand[7], T. Armstrong[8], H. Ashkar[9], M. Backes[1,6], V. Baghmanyan[10], V. Barbosa Martins[11], A. Barnacka[12], M. Barnard[6], R. Batzofin[13], Y. Becherini[14], D. Berge[11], K. Bernlöhr[3], B. Bi[15], M. Böttcher[6], C. Boisson[16], J. Bolmont[17], M. de Bony de Lavergne[7], M. Breuhaus[3], R. Brose[2], F. Brun[9], T. Bulik[18], T. Bylund[14], F. Cangemi[17], S. Caroff[17], S. Casanova[10], J. Catalano[19], P. Chambery[20], T. Chand[6], A. Chen[13], G. Cotter[8], M. Curyło[18], H. Dalgleish[1], J. Damascene Mbarubucyeye[11], I.D. Davids[1], J. Davies[8], J. Devin[20], A. Djannati-Ataï[21], A. Dmytriiev[16], A. Donath[3], V. Doroshenko[15], L. Dreyer[6], L. Du Plessis[6], C. Duffy[22], K. Egberts[23], S. Einecke[24], J.-P. Ernenwein[5], S. Fegan[25], K. Feijen[24], A. Fiasson[7], G. Fichet de Clairfontaine[16], G. Fontaine[25], F. Lott[1], M. Füßling[11], S. Funk[19], S. Gabici[21], Y.A. Gallant[26], G. Giavitto[11], L. Giunti[21,9], D. Glawion[19], J.F. Glicenstein[9], M.-H. Grondin[20], S. Hattingh[6], M. Haupt[11], G. Hermann[3], J.A. Hinton[3], W. Hofmann[3], C. Hoischen[23], T. L. Holch[11], M. Holler[27], D. Horns[28], Zhiqiu Huang[3], D. Huber[27], M. Hörbe[8], M. Jamrozy[12], F. Jankowsky[29], V. Joshi[19], I. Jung-Richardt[19], E. Kasai[1], K. Katarzyński[30], U. Katz[19], D. Khangulyan[31], B. Khélifi[21], S. Klepser[11], W. Kluźniak[32], Nu. Komin[13], R. Konno[11], K. Kosack[9], D. Kostunin[11], M. Kreter[6], G. Kukec Mezek[14], A. Kundu[6], G. Lamanna[7], S. Le Stum[5], A. Lemière[21], M. Lemoine-Goumard[20], J.-P. Lenain[17], F. Leuschner[15], C. Levy[17], T. Lohse[33], A. Luashvili[16], I. Lypova[29], J. Mackey[2], J. Majumdar[11], D. Malyshev[15], D. Malyshev[19], V. Marandon[3], P. Marchegiani[13], A. Marcowith[26], A. Mares[20], G. Martí-Devesa[27], R. Marx[29], G. Maurin[7], P.J. Meintjes[34], M. Meyer[19], A. Mitchell[3], R. Moderski[32], L. Mohrmann[19], A. Montanari[9], C. Moore[22], P. Morris[8], E. Moulin[9], J. Muller[25], T. Murach[11], K. Nakashima[19], M. de Naurois[25], A. Nayerhoda[10], H. Ndiyavala[6], J. Niemiec[10], A. Priyana Noel[12], P. O'Brien[22], L. Oberholzer[6], S. Ohm[11], L. Olivera-Nieto[3], E. de Ona Wilhelmi[11], M. Ostrowski[12], S. Panny[27], M. Panter[3], R.D. Parsons[33], G. Peron[3], S. Pita[21], V. Poireau[7], D.A. Prokhorov[35], H. Prokoph[11], G. Pühlhofer[15], M. Punch[21,14], A. Quirrenbach[29], P. Reichherzer[9], A. Reimer[27], O. Reimer[27], Q. Remy[3], M. Renaud[26], B. Reville[3], F. Rieger[3], C. Romoli[3], G. Rowell[24], B. Rudak[32], H. Rueda Ricarte[9], E. Ruiz-Velasco[3], V. Sahakian[36], S. Sailer[3], H. Salzmann[15], D.A. Sanchez[7], A. Santangelo[15], M. Sasaki[19], J. Schäfer[19], H.M. Schutte[6], U. Schwanke[33], F. Schüssler[9], M. Senniappan[14], A.S. Seyffert[6], J.N.S. Shapopi[1], K. Shiningayamwe[1], R. Simoni[35], A. Sinha[26], H. Sol[16], H. Spackman[8], A. Specovius[19], S. Spencer[8], M. Spir-Jacob[21], Ł. Stawarz[12], R. Steenkamp[1], C. Stegmann[23,11], S. Steinmassl[3], C. Steppa[23], L. Sun[35], T. Takahashi[37], T. Tanaka[38], T. Tavernier[9], A.M. Taylor[11], R. Terrier[21], J. H.E. Thiersen[6], C. Thorpe-Morgan[15], M. Tluczykont[28], L. Tomankova[19], M. Tsirou[3], N. Tsuji[39], R. Tuffs[3], Y. Uchiyama[31], D.J. van der Walt[6], C. van Eldik[19], C. van Rensburg[1], B. van Soelen[34], G. Vasileiadis[26], J. Veh[19], C. Venter[6], P. Vincent[17], J. Vink[35], H.J. Völk[3], S.J. Wagner[29], J. Watson[8], F. Werner[3], R. White[3], A. Wierzcholska[10], Yu Wun Wong[19], H. Yassin[6], A. Yusafzai[19], M. Zacharias[16], R. Zanin[3], D. Zargaryan[2,4], A.A. Zdziarski[32], A. Zech[16], S.J. Zhu[11], A. Zmija[19], S. Zouari[21] and N. Żywucka[6].

[1]University of Namibia, Department of Physics, Private Bag 13301, Windhoek 10005, Namibia
[2]Dublin Institute for Advanced Studies, 31 Fitzwilliam Place, Dublin 2, Ireland
[3]Max-Planck-Institut für Kernphysik, P.O. Box 103980, D 69029 Heidelberg, Germany
[4]High Energy Astrophysics Laboratory, RAU, 123 Hovsep Emin St Yerevan 0051, Armenia
[5]Aix Marseille Université, CNRS/IN2P3, CPPM, Marseille, France
[6]Centre for Space Research, North-West University, Potchefstroom 2520, South Africa
[7]Laboratoire d'Annecy de Physique des Particules, Univ. Grenoble Alpes, Univ. Savoie Mont Blanc, CNRS, LAPP, 74000 Annecy, France
[8]University of Oxford, Department of Physics, Denys Wilkinson Building, Keble Road, Oxford OX1 3RH, UK
[9]IRFU, CEA, Université Paris-Saclay, F-91191 Gif-sur-Yvette, France
[10]Instytut Fizyki Ja¸drowej PAN, ul. Radzikowskiego 152, 31-342 Kraków, Poland
[11]DESY, D-15738 Zeuthen, Germany
[12]Obserwatorium Astronomiczne, Uniwersytet Jagiello ński, ul. Orla 171, 30-244 Kraków, Poland
[13]School of Physics, University of the Witwatersrand, 1 Jan Smuts Avenue, Braamfontein, Johannesburg, 2050 South Africa
[14]Department of Physics and Electrical Engineering, Linnaeus University, 351 95 Växjö, Sweden
[15]Institut für Astronomie und Astrophysik, Universität Tübingen, Sand 1, D 72076 Tübingen, Germany
[16]Laboratoire Univers et Théories, Observatoire de Paris, Université PSL, CNRS, Université de Paris, 92190 Meudon, France
[17]Sorbonne Université, Université Paris Diderot, Sorbonne Paris Cité, CNRS/IN2P3, Laboratoire de Physique Nucléaire et de Hautes Energies, LPNHE, 4 Place Jussieu, F-75252 Paris, France
[18]Astronomical Observatory, The University of Warsaw, Al. Ujazdowskie 4, 00-478 Warsaw, Poland
[19]Friedrich-Alexander-Universität Erlangen-Nürnberg, Erlangen Centre for Astroparticle Physics, Erwin-Rommel-Str. 1, D 91058 Erlangen, Germany
[20]Université Bordeaux, CNRS/IN2P3, Centre d'Études Nucléaires de Bordeaux Gradignan, 33175 Gradignan, France
[21]Université de Paris, CNRS, Astroparticule et Cosmologie, F-75013 Paris, France
[22]Department of Physics and Astronomy, The University of Leicester, University Road, Leicester, LE1 7RH, United Kingdom
[23]Institut für Physik und Astronomie, Universität Potsdam, Karl-Liebknecht-Strasse 24/25, D 14476 Potsdam, Germany
[24]School of Physical Sciences, University of Adelaide, Adelaide 5005, Australia
[25]Laboratoire Leprince-Ringuet, École Polytechnique, CNRS, Institut Polytechnique de Paris, F-91128 Palaiseau, France
[26]Laboratoire Univers et Particules de Montpellier, Université Montpellier, CNRS/IN2P3, CC 72, Place Eugène Bataillon, F-34095 Montpellier Cedex 5, France
[27]Institut für Astro- und Teilchenphysik, Leopold-Franzens-Universität Innsbruck, A-6020 Innsbruck, Austria
[28]Universität Hamburg, Institut für Experimentalphysik, Luruper Chaussee 149, D 22761 Hamburg, Germany

[29]Landessternwarte, Universität Heidelberg, Königstuhl, D 69117 Heidelberg, Germany
[30]Institute of Astronomy, Faculty of Physics, Astronomy and Informatics, Nicolaus Copernicus University, Grudziadzka 5, 87-100 Torun, Poland
[31]Department of Physics, Rikkyo University, 3-34-1 Nishi-Ikebukuro, Toshima-ku, Tokyo 171-8501, Japan
[32]Nicolaus Copernicus Astronomical Center, Polish Academy of Sciences, ul. Bartycka 18, 00-716 Warsaw, Poland
[33]Institut für Physik, Humboldt-Universität zu Berlin, Newtonstr. 15, D 12489 Berlin, Germany
[34]Department of Physics, University of the Free State, PO Box 339, Bloemfontein 9300, South Africa
[35]GRAPPA, Anton Pannekoek Institute for Astronomy, University of Amsterdam, Science Park 904, 1098 XH Amsterdam, The Netherlands
[36]Yerevan Physics Institute, 2 Alikhanian Brothers St., 375036 Yerevan, Armenia
[37]Kavli Institute for the Physics and Mathematics of the Universe (WPI), The University of Tokyo Institutes for Advanced Study (UTIAS), The University of Tokyo, 5-1-5 Kashiwa-no-Ha, Kashiwa, Chiba, 277-8583, Japan
[38]Department of Physics, Konan University, 8-9-1 Okamoto, Higashinada, Kobe, Hyogo 658-8501, Japan
[39]RIKEN, 2-1 Hirosawa, Wako, Saitama 351-0198, Japan